# Photonic bandgap plasmonic waveguides

Andrey Markov,[1] Carsten Reinhardt,[2] Bora Ung,[1] Andrey B. Evlyukhin,[2] Wei Cheng,[2] Boris N. Chichkov[2] and Maksim Skorobogatiy[1,*]

[1]*École Polytechnique de Montréal, Génie Physique, Québec, Canada*
[2]*Laser Zentrum Hannover e.V., Hannover, Germany*
*\*Corresponding author:* [maksim.skorobogatiy@polymtl.ca](maksim.skorobogatiy@polymtl.ca)

A novel type of a plasmonic waveguide has been proposed featuring an "open" design that is easy to manufacture, simple to excite and that offers a convenient access to a plasmonic mode. Optical properties of photonic bandgap (PBG) plasmonic waveguides are investigated experimentally by leakage radiation microscopy and numerically using the finite element method confirming photonic bandgap guidance in a broad spectral range. Propagation and localization characteristics of a PBG plasmonic waveguide have been discussed as a function of the wavelength of operation, waveguide core size and the number of ridges in the periodic reflector for fundamental and higher order plasmonic modes of the waveguide.

Surface plasmon resonance (SPR) is a prominent optical phenomenon, which involves the resonant excitation of plasmon or electromagnetic waves coupled to collective oscillations of free electrons in a metal, over a metal/dielectric interface. Recently, SPR-based sensors have attracted significant attention due to their record sensitivities and low noise levels. Propagating at the metal/dielectric interface, surface plasmons are highly sensitive to changes in the refractive index of a dielectric. This feature constitutes the core of many SPR sensors.

Typically, these sensors are implemented in the Kretschmann-Raether prism geometry to direct p-polarized light through a glass prism and reflect it from a thin metal (Au, Ag) film deposited on the prism facet [1]. The presence of a prism allows phase matching of an incident electromagnetic wave with a plasmonic wave at the metal/ambient dielectric interface at a specific combination of the angle of incidence and wavelength, which is resonantly dependent on the refractive index of an ambient medium. Mathematically, the phase matching condition is expressed as an equality between a plasmon wave vector and a projection of a wave vector of an incident wave along the interface. Since the plasmon excitation condition depends resonantly on the value of the refractive index of an ambient medium within 100 - 300 nm from the interface, the method enables detection of biological binding events on the metal surface with unprecedented sensitivity [2]. In SPR biosensors, refractive index changes due to biological reactions are controlled by monitoring angular [2], spectral [3] or phase characteristics [4,5] of the reflected light. However, the high cost and large size of commercially available systems make them useful only in a laboratory, while many important field and applications remain out of the reach for this method.

Using optical waveguides and fibers instead of a bulk prism configuration in plasmonic sensors offers miniaturization, a high degree of integration and remote sensing capabilities. In such sensors, one launches the light into a waveguide core and then uses coupling of a guided mode to a plasmonic mode to probe for the changes in the ambient environment. In the majority of fiber implementations (with an exception of microstructured fibers), one typically strips the fiber polymer jacket and polishes off the fiber cladding until the fiber core is exposed. Then, a metal layer is deposited directly onto the fiber core and the functionalized surface of the fiber core is exposed to an analyte. To excite efficiently a surface plasmon at the metal/analyte interface the phase matching condition between the plasmon and waveguide modes has to be satisfied, which mathematically amounts to the equality between their modal propagation constants. Near the point of phase matching, most of the energy launched into a waveguide core mode should be efficiently transferred into a plasmon mode [6].

Such an approach based on planar waveguides has been indeed demonstrated in the visible to provide several compact designs of SPR biosensors [6-8]. However, the phase matching between the plasmon mode and the fundamental waveguide mode is not easy to realize. This is related to the fact that the effective refractive index of such a mode is close to the refractive index of the core material, which is typically larger than $n = 1.45$ due to the materials limitations. The refractive index of a plasmon is close to the refractive index of the ambient medium which is typically air with $n = 1$ or water, having $n = 1.3$. Thus, a large discrepancy in the effective refractive indices makes it hard to achieve phase matching between the two modes, with an exception of high frequencies ($\lambda < 650$ nm) where the plasmon dispersion relation deviates substantially from that of an analyzed material. Thus, due to practical limitation on the lowest value of the waveguide core and cladding refractive indices, single mode TIR waveguide-based sensors were demonstrated almost exclusively in the visible where phase matching condition is easier to enforce.

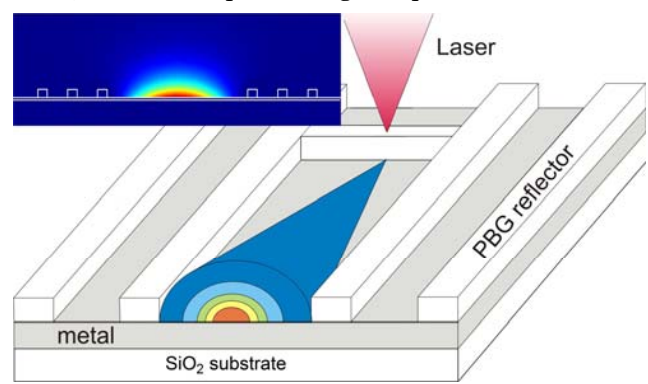

Fig. 1. (Color online) Photonic bandgap plasmonic waveguide.

In this article, we propose a novel type of plasmonic waveguide featuring an "open" design that is highly integrated, easy to manufacture, simple to excite and that offers a convenient access to the plasmonic mode. We believe that the main utility of this type of waveguides will be in the design of integrated plasmonic sensors.

The geometry of the proposed waveguide is illustrated in Fig. 1. The plasmonic mode propagates on the surface of a thin metallic film allocated on a glass slide. The metallic film used in numerical simulations and the experimental investigations consists of a layer of gold with 50 nm thickness. Surface plasmon polaritons (SPP) may be excited inside the waveguiding structure by Kretschmann or grating/ridge coupling. The surface plasmon is confined in the transverse plane direction by a periodic sequence of ridges which constitute a PBG reflector. Due to its design the waveguide on the one hand offers a convenient access to its metallic surface, while on the other hand it allows transverse confinement of the plasmonic mode which is necessary for a compact sensor design. Finally the design is very simple for manufacturing using either lithographic or direct writing techniques (two-photon polymerization).

The width of the core region should be chosen several times larger than the wavelength in order to support at least the fundamental mode. In our numerical simulations the width was 7.5 μm. For the sake of simplicity of presentation we consider the core region to have the height equal to that of the ridges which is 600 nm. The width of the ridges was at first estimated by analogy with Bragg fibers and planar multilayer stacks [9]. Particularly, in these systems the bandgap at a certain wavelength λ is the widest when the quarter wave

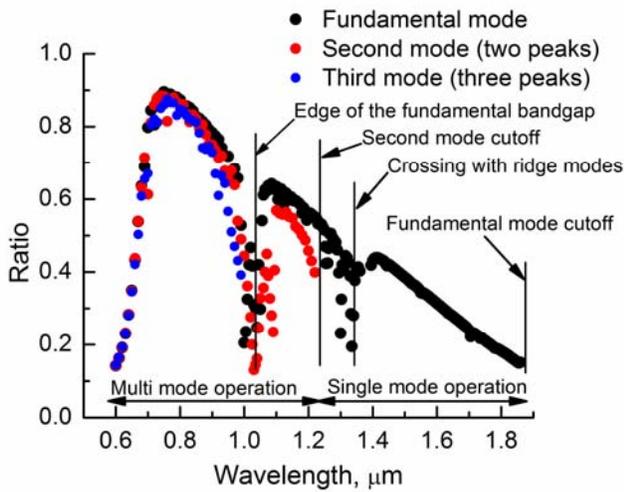

Fig. 2. (Color online) Ratio of power guided inside the waveguide core to the total modal power for the three lowest order modes

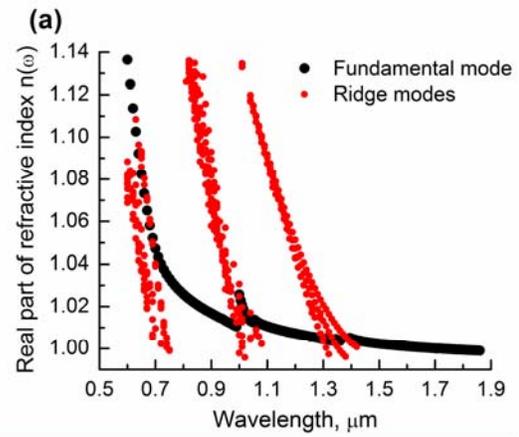

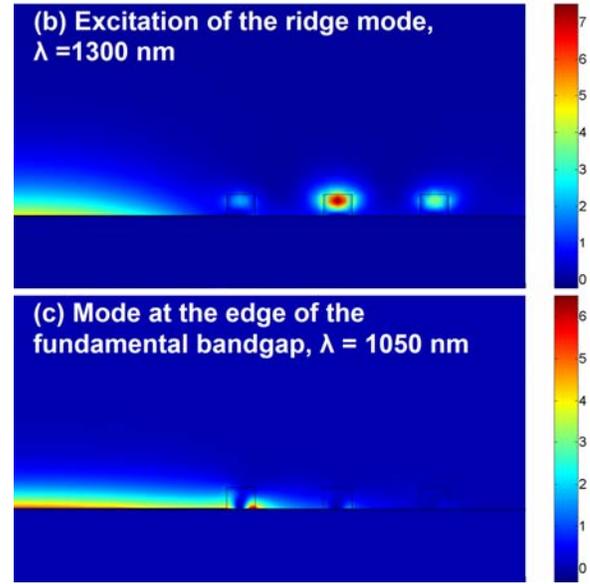

Fig. 3. (Color online) (a) Real part of the refractive index of the fundamental mode and the ridge modes. Transverse electric field distributions for the (b) fundamental mode exhibiting avoiding crossing with the ridge modes, (c) fundamental mode in the vicinity of a bandgap edge.

resonant condition for the layer thickness is satisfied [10]:

$$d_0 = \lambda \Big/ \left(4\sqrt{n_{ridge}^2 - n_{eff}^2}\right) \quad (1)$$

where $n_{eff}$ is close to 1. This equation gives the minimum value of thickness around 370 nm for λ=1550 nm. Alternatively, one can use larger ridge thickness $d_0 \cdot (2k+1)$ where k is an integer number. However, for larger ridge sizes, there is a risk of drawing in the modes localized at the ridges themselves, which would lead to fracturing the fundamental bandgap. After numerical optimizations and from practical considerations the ridge width was chosen to be 500 nm with inter-ridge distance 1200 nm.

Detailed behavior of the first three plasmonic modes having their maximum intensity in the core center was then studied numerically. Two effects significantly affecting the SPP propagation have been observed. First, it is revealed that at certain resonant frequencies (in the range λ = 1300 – 1400 nm in Fig. 2), the fundamental core-guided plasmonic mode exhibits crossing with the ridge modes leading to a significant modal field presence inside the reflector ridges (for the electric field distribution in this case see Fig. 3(b)). The dependence of the refractive index of the ridge modes as a function of wavelength is depicted in Fig. 3 (a). There is one more wavelength region near 1000 nm where the fundamental mode and ridge modes are simultaneously excited. Near the wavelength of the second dip in the transmitted power at $\lambda = 1050$ nm the mode shows both the avoiding crossing with the ridge modes and delocalization due to proximity to the edge of a fundamental bandgap (see Fig. 3(c)).

We also observe that the fundamental plasmonic mode has a high wavelength cutoff near 1800 nm above which there is no core guidance. The propagation loss of a SPP mode due to absorption in the metal layer and leakage of light into the substrate have also been investigated. As operation frequency decreases, so does the propagation loss of a SPP mode which is primarily due to higher absorption loss of the gold film at higher frequencies.

Second and higher order plasmonic modes of the waveguide have lower values of cutoff wavelengths separating single-mode and multi-mode regimes of the waveguide (Fig. 2). For the wavelengths between 1200 and 1800 nm the plasmonic waveguide provides a single-mode guiding regime and for even shorter wavelengths several modes exist in the waveguide with approximately the same absorption losses.

Finally, we report preliminary experimental investigation of the PBG waveguides. The ridges are fabricated by two-photon polymerization of spin-coatable epoxy-based resist mr-NIL 6000.5 (micro resist technology GmbH) on a gold covered glass slide [11]. The gold film is fabricated by thermal evaporation in high vacuum of $10^{-6}$ mbar. SPPs at different wavelengths are excited by focusing the linearly polarized laser radiation onto an additional ridge placed perpendicular to the main system of ridges which constitute the bandgap structure, see again Fig. 1. To visualize the plasmon modes experimentally we have used a leakage radiation imaging setup [12], together with an NIR InGaAs camera for detection in the range of 1100 nm and 1700 nm (Hamamatsu C10633-23).

The PBG waveguide investigated experimentally was designed for wavelength $\lambda = 1550$ nm with the geometric parameters identical to those used in numerical simulations. At first, the properties of the waveguide were examined at $\lambda = 1550$ nm (see Fig. 4(a)). Experimental results agree with numerical calculations confirming single-mode guiding at this wavelength. Propagation characteristics of the waveguide were also investigated at the lower wavelength $\lambda = 974$ nm. Numerical simulations predicted effectively two modes operation of the waveguide for this wavelength (third mode is close to its cutoff frequency). In our experiments we have indeed observed simultaneous excitation of the two low order modes leading to a mode beating and an oscillation of the guided intensity (see Fig. 4(b)). From the experiment we conclude that the excited SPPs are localized and guided inside of the hollow waveguide core and can be associated with the low order modes of the corresponding photonic bandgap plasmonic waveguide. Core guided modes have a predominant field presence in the air core. The propagation length, defined as the 1/e intensity decay of the SPP inside the waveguide, can be measured to 63 µm for an excitation wavelength of $\lambda = 1550$ nm. Due to the strong sensitive of the metal absorption to the wavelength of operation the propagation length at $\lambda = 974$ nm is reduced approximately 3 times to about 20 µm.

In conclusion, we have proposed a novel type of plasmonic waveguide featuring an "open" design that is easy to manufacture, simple to excite and that offers a convenient access to a plasmonic mode. Potentially, this allows usage of this type of waveguides as integrated plasmonic sensors. Propagation and localization characteristics of the waveguide have been investigated numerically using the finite element method as a function of the wavelength of operation, waveguide core size. Finally, guidance at several wavelengths of interest was confirmed experimentally by leakage radiation spectroscopy for a particular PBG plasmonic waveguide design fabricated using two photon polymerization.

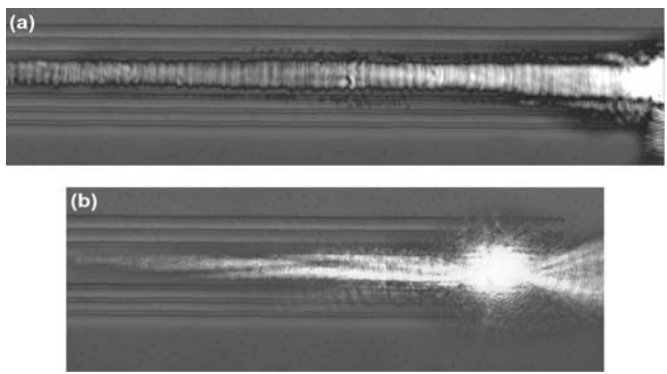

Fig. 4. (Color online) Photograph of a photonic bandgap plasmonic waveguide designed for 1550 nm. Overlaid is the modal intensity distribution inside the waveguide (a) 1550 nm (single mode guidance), (b) 974 nm (two mode guidance).